\documentstyle[12pt]{article}

\begin{document}
\noindent
\begin{center}

{\bf NUCLEAR MANY-BODY PROBLEM AT FINITE TEMPERATURE: A TFD APPROACH}


D.S.~Kosov$^1$, A.I.~Vdovin$^2$, J.~Wambach$^3$

$^1$
ITP, University of T\"ubingen,
D-72076, T\"ubingen, Germany

$^2$ BLTP,
Joint Institute for Nuclear Research,
141980 Dubna, Russia

$^3$
INP, Technical University of Darmstadt,
D-64289, Darmstadt, Germany

\end{center}

\begin{abstract}
Based on the formalism of thermo-field dynamics a new approach for
studying collective excitations in hot finite Fermi systems
is presented. Two approximations going beyond the thermal RPA
namely renormalized thermal RPA and thermal second RPA are formulated.
\end{abstract}

\section{Introduction}

A standard technique for treating  quantum many-body systems at finite
temperature $T$ is the thermal Green function (Matsubara)
method. We here outline an alternative approach,
based on the concepts of thermo-field dynamics (TFD) \cite{tfd1,tfd2},
to treat collective excitations of hot finite Fermi systems.
In this context TDF has two appealing features:
\begin{itemize}
\item[--]temperature effects arise explicitly as $T$-dependent vertices,
providing a good starting point for various approximations
\item[--]a generalization to the time-dependent situation is easy, since
temperature and time are independent variables in TFD (real-time formalism).
\end{itemize}

As will be detailed below, both features allow for straightforward extensions
of well-established zero-temperature approximations.
Some aspects of the TFD application to hot nuclei have been
discussed previously by Hatsuda~\cite{hats} and Tanabe~\cite{tan}.

\section{TFD: Basic Elements}

The extension of quantum field theory to finite
temperature requires a doubling of the field degrees of freedom.
In TFD this doubling  is achieved
by introducing an additional tilde space \cite{tfd1,tfd2}.
A tilde conjugate operator $\tilde A$ is assigned an operator
$A$ (acting in ordinary space) through the tilde conjugation rules
 $$ (AB)^{\sim} = \tilde A \tilde B; \;\;\;
 (a A + b B)^{\sim} = a^* \tilde A + b^* \tilde B \; ,
$$
where $A$ and $B$ represent ordinary operators and $a \: , b$ are c-numbers.
The asterisk denotes the complex conjugate. The tilde operation commutes
with hermitian conjugation and any tilde and non-tilde
operators are assumed to commute or anticommute with each other.
The thermal Hamiltonian ${\cal H} = H - \tilde H$ is the time-translation
operator and the excitation spectrum of the system is obtained
by the diagonalization of $\cal H$.

\noindent
The temperature-dependent vacuum $|\Psi_{0} (T) \rangle $ is the
eigenvector of ${\cal H}$ with eigenvalue~$0$:
$$ {\cal H}|\Psi_{0}(T)\rangle =0 $$

\noindent
If one determines the thermal vacuum state as
$$\Psi_{0} (T) \rangle = \frac{1}{\sqrt{Tr(exp(- H/T ))}} \sum_{n}
exp(-\frac{E_{n}}{2T})|n\rangle \otimes |\tilde{n}\rangle $$
where $E_{n}, |n\rangle$ and $|\tilde n\rangle$ are eigenvalues,
eigenvectors and their tilde counterparts of the Hamiltonian $H$,
respectively,
the expectation value $\langle \Psi_{0} (T)|O|\Psi_{0} (T) \rangle$
will exactly correspond to the grand canonical ensemble average
$\ll O \gg $ of a given observable~$O$.

\section{Temperature-dependent Fock space in HF}

Restricting ourselves to two-body interactions, the $N$-body
Hamiltonian is written~as

$$
H=\sum_{12} t_{12} a_{1}^{+} a_{2} +\frac{1}{4}
\sum_{1234} V_{1234} a_{1}^{+} a^{+}_{2} a_{4} a_{3} \;, \eqno(1)
$$
where $a^{+},a$ are fermion creation and annihilation operators.
The one-body part
$$
t_{12}=T_{12}- \lambda \delta_{12}
$$
contains the kinetic energy matrix $T_{12}$ as well as the chemical
potential $\lambda$.

Using the Wick theorem one can expand the Hamiltonian (1) in normal
order (:~...~:) with respect to the temperature-dependent ground state
$|\Psi_{0}(T) \rangle $ and it has the following form:

$$
H=h_{0} +  h_{11} +h_{22} \; , \eqno(2)
$$
where
$$
h_{0}=\sum_{12} t_{12} \rho_{21} + \frac{1}{2} \sum_{1234}
V_{1234} \rho_{31} \rho_{42} \; ,
$$

$$
h_{11}= : \sum_{12} t_{12} a_{1}^{+} a_{2}
+ \sum_{1234} V_{1234} \rho_{42} a_{1}^{+}a_{3}: \; ,
$$

$$
h_{22}= :\frac{1}{4}\sum_{1234} V_{1234} a_{1}^{+} a_{2}^{+} a_{4} a_{3}:
\; .
$$

\noindent
We denote the density matrix as

$$
\rho_{ij} =\langle \Psi_{0} (T) | a^{+}_{j} a_{i} | \Psi_{0} (T) \rangle \; .
$$

\noindent
To diagonalize the $h_{11}$-part of $H$ which is
quadratic in $a^{+} , a$ one has to perform the unitary transformation
\cite{blaizot}:
$$
a^{+}_{l} =\sum_{k} D^{*}_{lk} \alpha^{+}_{k} \; ,
a_{l} =\sum_{k}  D_{lk} \alpha_{k}   \; ,
$$
where $DD^{+}=D^{+}D=I$ and
 $\alpha^{+}, \alpha$ are new fermionic (``quasiparticle") operators.

Requiring  $h_{11}$  to be diagonal in $\alpha^{+}, \alpha$,
a system of equations for the matrix elements $D$ as well as
the single-quasiparticle energy $\varepsilon_{1}$ is obtained:
$$
\sum_{2} \left( t_{12} + \sum_{34} V_{1324} \rho_{43} \right)
D_{25} =\varepsilon_{5} D_{15}
\eqno(3)
$$
with the auxiliary condition

$$
\sum_{1} \rho_{11} = N \; .
\eqno(4)
$$
which implies conservation of the average number of particles.
In the density matrix
$$
\rho_{12} =\langle \Psi_0 (T)| a_{2}^{+} a_{1} |\Psi_0 (T) \rangle =
\sum_{k} D^{*}_{2k} D_{1k} n_k
$$
the quasiparticle occupation numbers $n_k= \langle \Psi_0 (T)|
\alpha_{k}^{+} \alpha_{l} |\Psi_0 (T) \rangle \delta_{kl}$
can be determined once an explicit expression of the thermal vacuum
is specified.

In practice it is impossible to find the exact thermal vacuum for
the full Hamiltonian (2). In setting up approximation
schemes the usual starting point is thermal mean-field theory
(the thermal Hartree-Fock approximation (THF)). In this case
the thermal vacuum is an eigenvector of the
uncorrelated thermal Hamiltonian :
$$
(h_{11}-\tilde{h}_{11})|0(T)\rangle =\sum_{i}\varepsilon_{i}(\alpha^{+}_{i}
\alpha_{i}
-\tilde{\alpha}^{+}_{i} \tilde{\alpha}_{i}) |0(T)\rangle  = 0 \eqno(5)
$$

\noindent
The solutions  of eq.~(5) define the vacuum $|0(T)\rangle$ for so-called
thermal
quasiparticles $\beta,\tilde{\beta}$:
$$ \beta_{i} =
x_{i}\alpha_{i} - y_{i}\tilde{\alpha}^{+}_{i} $$
$$\tilde{\beta}_{i} =
x_{i}\tilde{\alpha}_{i} + y_{i}\alpha^{+}_{i} $$
$$ \beta_{i} |0(T)\rangle =\tilde{\beta}_{i} |0(T)\rangle = 0
$$
where the coefficients $x_i , y_i$ denote the thermal Fermi
occupation probabilities of the states $\alpha^{+}_{i} |0\rangle$:
$$ x_{i} = \sqrt{1-f_{i} }\; , \; y_{i}= \sqrt{f_{i} }  $$

$$ f_{i} = \frac{1}{1+\exp(\varepsilon_{i}/T)}$$

\noindent
The chemical potential $\lambda$ is adjusted
to fulfill eq.~(4) ensuring that
the average number of particles in the system remains fixed.

In terms of the quasiparticle operators $\alpha^{+}, \alpha$
the THF-vacuum state
$$
|0(T)\rangle = \prod_{i} \left( x_{i} + y_{i}\alpha^{+}
\tilde{\alpha}^{+} \right)|0\rangle |\tilde{0}\rangle
$$
is similar to the BCS-vacuum for systems with pairing.

Based on the mean field we are now able to construct a temperature-dependent
Fock space. The Fock space is generated by acting with thermal
quasiparticle creation operators $\beta^{+}, \tilde{\beta}^{+}$ on the
thermal vacuum
$$
\beta^{+}|0(T)\rangle, \tilde{\beta}^{+}|0(T)\rangle,
\beta^{+} \tilde{\beta}^{+}|0(T)\rangle, ...
$$

\noindent
After the transformation to thermal quasiparticles the thermal Hamiltonian
${\cal H}$  takes the form:
$$
{\cal H} = {\cal H}_{11} +{\cal H}_{22} +{\cal H}_{40}
+{\cal H}_{04}+{\cal H}_{31}
+{\cal H}_{13}                      \eqno(6)
$$
where  $ {\cal H}_{mn}$ consists of terms of the type
$ (\beta^{+})^{m} (\beta)^{n}$.
\begin{eqnarray}
{\cal H}_{11}& =&\frac{}{}\sum_{1} \varepsilon_{1} (\beta^{+}_{1}
\beta_{1} -\tilde{\beta}^{+}_{1} \tilde{\beta}_{1}) \nonumber\\
{\cal H}_{22}&=&
\frac{1}{4}\sum_{1234} U_{1234} (x_1 x_2 x_3 x_4 -
y_1 y_2 y_3 y_4) (\beta_{1}^{+} \beta_{2}^{+} \beta_{4} \beta_{3}
-\tilde{\beta}^{+}_{1} \tilde{\beta}^{+}_{2} \tilde{\beta}_{4}
 \tilde{\beta}_{3}) +
\nonumber\\
&&+ \frac{}{}\sum_{1234} U_{1234} (x_1 y_2 y_3 x_4 - y_1 x_2 x_3 y_4)
\beta_{1}^{+} \tilde{\beta}^{+}_{3}\tilde{\beta}_{2}
 \beta_{4}
\nonumber\\
{\cal H}_{04}&=&
\frac{1}{4} \sum_{1234} U_{1234}
(x_1 x_2 y_{3} y_{4} - y_1 y_2 x_3 x_4)\beta_{1}\beta_{2} \tilde{\beta}_{4}
 \tilde{\beta}_{3}
\nonumber\\
{\cal H}_{13}&=&
\frac{1}{2} \sum_{1234} U_{1234}(x_1 x_2 x_3 y_4 - y_1 y_2 y_3 x_4)
(\beta^{+}_{3} \tilde{\beta}_{4}\beta_{2} \beta_{1}-
\tilde{\beta}^{+}_{3}\tilde{\beta}_{1}\tilde{\beta}_{2}\beta_{4})
\nonumber
\end{eqnarray}
$$
{\cal H}_{31}={\cal H}_{13}^{+}\, , \;\;
{\cal H}_{40}={\cal H}_{04}^{+}
$$
where
 $$
 U_{1234} =\sum_{5678} V_{5678} D^{*}_{51} D^{*}_{62} D_{73} D_{84}.
 $$

\noindent
Given the Fock space the solution of any statistical problem is thus reduced
to a diagonalization of the thermal Hamiltonian.

\section{TRPA}

Based on the above formulation, a straightforward way to solve the RPA
at finite temperature (TRPA) is to apply the equation of
motion method \cite{rowe}. As for $T=0$, one has a Raleigh-Ritz variational
principle with the thermal Hamiltonian (6):
$$ \langle \Psi_{0}(T)|\left[ \delta Q_{\nu}, [{\cal H}, Q^{+}_{\nu}]
\right]| \Psi_{0}(T)\rangle =
\omega_{\nu} \langle \Psi_{0}(T)|\left[ \delta Q_{\nu},
 Q^{+}_{\nu} \right]| \Psi_{0}(T)\rangle \eqno(7)
$$

The potentially exact statistical variational problem (7)
cannot be solved in practice and the class of variational functions has
to be restricted.
The simplest approximation amounts to the following trial wave function
(phonon wave function):
$$ Q^{+}_{\nu} =\sum_{12} \psi^{\nu}_{12} \beta^{+}_{1} \tilde{\beta}^{+}_{2}-
\phi^{\nu}_{12} \tilde{\beta}_{2} \beta_{1} =
\sum_{12} \psi^{\nu}_{12} A^{+}_{12} - \phi^{\nu}_{12} A_{12}
\eqno(8)
$$
If, in addition, the exact thermal vacuum in (7) is replaced by thermal
HF-vacuum, the exact matrix equation (7) reduces to the usual
TRPA equations for the amplitudes
$\psi^{\nu}_{24} , \phi^{\nu}_{24}$ and the excitation energy $\omega_{\nu}$:
$$
\varepsilon_{24}\psi^{\nu}_{24} +
\frac{1}{2} x_2 y_4 \sum_{13}U_{1234}
 (x_1 y_3 \phi^{\nu}_{13} + x_3 y_1 \psi^{\nu}_{31})
$$
$$
\quad-\frac{1}{2} x_4 y_2 \sum_{13}U_{1234}
 (x_3 y_1 \phi^{\nu}_{13} + x_1 y_3 \psi^{\nu}_{31}) = \omega_{\nu}
\psi^{\nu}_{24}
$$
$$
\varepsilon_{42}\phi^{\nu}_{42} +
\frac{1}{2} x_4 y_2 \sum_{13}U_{1234}
 (x_1 y_3 \phi^{\nu}_{13} + x_3 y_1 \psi^{\nu}_{31})
$$
$$
\quad-\frac{1}{2} x_2 y_4 \sum_{13}U_{1234}
 (x_3 y_1 \phi^{\nu}_{13} + x_1 y_3 \psi^{\nu}_{31}) = -\omega_{\nu}
\phi^{\nu}_{42}
 \; .
$$
At this stage only the terms
${\cal H}_{11}$, ${\cal H}_{22}$, ${\cal H}_{40}$ and ${\cal H}_{04}$
play a role and this part of $\cal H$ becomes diagonal in the
phonon operators.
$$
{\cal H} = \sum_{\nu} \omega_{\nu}  Q^{+}_{\nu}Q_{\nu} +{\cal H}_{13} +
{\cal H}_{31}
\; .
$$
In the TRPA the $Q^{+}, Q$ operators are
boson operators (the same holds for the bifermionic operators
$A^{+}_{12} \mbox{~and~} A_{12}$)
The index $\nu$ runs over TRPA solutions with positive and negative
energies $\omega_{\nu}$.  The negative-energy solutions appear
naturally and ensure that the set of thermal
phonon states is complete within TRPA approximation:
$$
I=\sum_{\nu}\sum_{k =0}^{\infty}
 (Q^{+}_{\nu})^k| \Psi_{0}(T)\rangle \langle \Psi_{0}(T)|(Q_{\nu})^k
\; ,
$$
where $| \Psi_{0}(T)\rangle$ is the vacuum for thermal phonons,
i.e. $Q_{\nu}| \Psi_{0}(T)\rangle$ = 0. In terms of thermal quasiparticle
operators, the TRPA vacuum is given by

$$
\Psi_{0} (T) = \frac{1}{\sqrt N} \prod_{\nu}
\exp \frac{1}{4}
\left[ \sum_{1234}(\psi^{-1})^{\nu}_{12}\phi^{\nu}_{34}
\beta^{+}_{1}\tilde\beta^{+}_{2} \beta^{+}_{3}\tilde\beta^{+}_{4}\right]
|0(T)\rangle \; .
$$

\section{Beyond TRPA}
The derivation of the TRPA, presented above, suggests that
improvements known at zero temperature can be taken over straightforwardly.
Below we demonstrate this for the so-called renormalized RPA
[6-9] as
well as the second RPA \cite{srpa,koswam}.

Within the renormalized RPA, the bifermionic operators
$A^{+}_{12}$ and $A_{12}$ are no longer bosonic, as in RPA,
but obey modified commutation relations:
$$
\left[A_{12} , A^{+}_{34} \right]
= \delta_{13} \delta_{24} \left(1 - q_1 - q_2 \right) \;,
$$
where $q_i$ are c-numbers. At finite temperature these are defined as
follows:
$$
\langle \Psi_{0} (T) |\beta^{+}_{1} \beta_{2}| \Psi_{0} (T) \rangle =
\langle \Psi_{0} (T)| \tilde{\beta}^{+}_{1} \tilde{\beta}_{2}
|\Psi_{0} (T) \rangle = \delta_{12} q_{1} \;.
$$
The new thermal vacuum state  $|\Psi_{0} (T)\rangle $ is different from
the Hartree-Fock or TRPA vacuum state and will be determined below.

\noindent
The substitution~\cite{rowe,ken}
$$
b_{12}= \frac{A_{12}}{\sqrt{1-q_{12}}}\;, \quad
b^{+}_{12}= \frac{A^{+}_{12}}{\sqrt{1-q_{12}}} \; ,
$$
with $q_{12}= q_{1}+ q_{2}$, leads to bosonic commutation rules for the
operators $b_{12} , b^{+}_{12}$. In analogy to the TRPA, we introduce phonon
creation operators as
$$ Q^{+}_{\nu} =
\sum_{12} \psi^{\nu}_{12} b^{+}_{12} - \phi^{\nu}_{12} b_{12}
\eqno(9)$$
and define the thermal vacuum state as the vacuum for the new phonon
operators.
Then
$$
|\Psi_0(T) \rangle = \frac{1}{\sqrt{N}}  \exp{S}
| 0(T)\rangle
$$
$$
S=\frac{1}{2}
\sum_{1234} C_{1234} b^{+}_{12} b^{+}_{34}
$$
where the matrix $C_{1234}$ is defined through:
$$ \sum_{12} \psi^{\nu}_{12} C_{1234}=\phi^{\nu}_{34}\; .
$$
For practical applications it is more appropriate to write
the system of equations of the
renormalized TRPA in terms of new variables, namely
 $$ \Psi^{\nu}_{24} =\frac{1}{x_{2}y_{4} +y_{2}x_{4}} (\psi^{\nu}_{24}-
 \phi_{42}^{\nu}) \qquad , \qquad
  \Phi^{\nu}_{24} =\frac{1}{x_{2}y_{4} -y_{2}x_{4}} (\psi^{\nu}_{24}+
 \phi_{42}^{\nu})\; .
 $$
The result is
$$
\sum_{2} \left[ t_{12} + \sum_{345} V_{1324} D^{*}_{26} D_{46}
( x^{2}_{5} q_{5} +y_{5}^{2} (1-q_{5})) \right]
D_{26} =\varepsilon_{6} D_{16} \; ,
\eqno(10)
$$
$$
\sum_{12} D^{*}_{12} D_{12}( x^{2}_{2} q_{2} +y_{2}^{2} (1-q_{2}))  = N \; ,
\eqno(11)
$$
$$
\varepsilon_{24} \Psi^{\nu}_{24}+
\frac{1}{2}\sum_{13}\sum_{5678}
\left( V_{5678} D^{*}_{51} D^{*}_{62} D_{73} D_{84} \right)
\sqrt{1-q_{13}}\sqrt{1-q_{24}}(n_1 -n_3)
\Psi^{\nu}_{31}
 =\omega_{\nu} \Psi_{24}^{\nu}
\eqno(12)
$$
$$
q_{1}= \sum_{\nu >0 \atop 2} \left[ (\Psi^{\nu}_{12}
x_2 y_1 )^{2} + (\Psi^{\nu}_{21} y_2 x_1)^{2} \right] \; .
\eqno(13)
$$

\noindent
This system of equations is nonlinear. The important feature
is that the  mean-field variables not only dependent on $T$ but also on
the collective variables $\Psi , \Phi$. This coupling disappears in TRPA.
If $q_i$=0, the system splits into two independent parts: eqs.~(10)-(11) and
eq.~(12). The latter then corresponds to the TRPA equation.

Only the harmonic part of the thermal Hamiltonian
is taken into account within the TRPA and renormalized TRPA.
To incorporate effects of ${\cal H}_{31}$ and ${\cal H}_{13}$
one has to extend the class of variational functions by including
thermal four-quasiparticle configurations~\cite{koswam}:

$$ Q^{+}_{\nu} =\sum_{12} X^{\nu}_{12} \beta^{+}_{1} \tilde{\beta}^{+}_{2}-
Y^{\nu}_{12} \tilde{\beta}_{2} \beta_{1}
+ \sum_{1>2,3>4} X^{\nu}_{1234} \beta^{+}_{1} \beta^{+}_{2}
\tilde{\beta}^{+}_{3} \tilde{\beta}^{+}_{4} -
 Y^{\nu}_{1234} \tilde{\beta}_{4} \tilde{\beta}_{3}
\beta_2 \beta_1
\eqno(14)
$$
Direct application of the variational principle to (14) yields
the thermal second RPA matrix equation:
$$
\left(
\begin{array}{cc}
A & B  \\
B^{*} & A^{*}
\end{array}
\right)
\left(
\begin{array}{c}
X \\
-Y
\end{array}
\right)         =
\omega_{\nu}
\left(
\begin{array}{c}
X \\
Y
\end{array}
\right)\; .
$$
The stability matrix is composed of $2\times2$-supermatrices:
$$
A=
\left(
\begin{array}{cc}
A_{12;34} & A_{12;3456} \\
A_{1234;56} & A_{1234;5678}
\end{array}
\right)
\qquad ;
\qquad
B=
\left(
\begin{array}{cc}
B_{12;34} & B_{12;3456} \\
B_{1234;56} & B_{1234;5678}
\end{array}
\right)
$$
and the amplitudes $X^{\nu}$ and $Y^{\nu}$ are two-component supervectors:
 $$
X^{\nu}=
\left(
\begin{array}{c}
X^{\nu}_{12} \\
X^{\nu}_{1234}
\end{array}
\right)
\qquad ;
\qquad
Y^{\nu}=
\left(
\begin{array}{c}
Y^{\nu}_{12} \\
Y^{\nu}_{1234}
\end{array}
\right)\; .
$$

\noindent
Explicitly, the supermatrices $A$ and $B$ are expressed as
 $\Psi_{0} (T)$-expectation
values of double commutators of the thermal Hamiltonian
$\cal H$  with different kinds of two and four thermal quasiparticle
operators.

If one approximates $|\Psi_{0} (T) \rangle$ by $|0 (T) \rangle$ (as in the
TRPA), the
matrix elements $A_{12;34}$ and $B_{12;34}$ are the usual TRPA matrix
elements while $A_{12;3456}$, $A_{1234;56}$, and $A_{1234;5678}$ describe
the coupling of two-quasiparticle states with more complex ones:
$$
\begin{array}{lcl}
A_{12;34}&=&( \varepsilon_3 -\varepsilon_4 ) \delta_{23} \delta_{14}
+ U_{2413} (x_1 x_4 y_2 y_3 - x_2 x_3 y_1 y_4) \\
&&\\
A_{12;3456}&=&a(65)
U_{3426} (x_3 x_4 y_2 y_6 - x_6 y_2 y_3 y_4) \delta_{15} +\\
&&\\
&&+a(34)U_{5613} (x_5 x_6 y_1 y_3 - x_3 y_1 y_5 y_6) \delta_{24} \\
\end{array}
$$
$$
\begin{array}{lcl}
A_{1234;5678}&=&\delta_{45} \delta_{36} \delta_{27} \delta_{18}
(\varepsilon_{3} +\varepsilon_{4} -\varepsilon_{1} -\varepsilon_{2})+\\
&&\\
&&+\delta_{27} \delta_{18} U_{4356} (x_{3} x_{4} x_{5} x_{6} -
y_{3} y_{4} y_{5} y_{6})- \\
&&\\
&&- \delta_{45} \delta_{36} U_{2178} (x_{1} x_{2} x_{7} x_{8} -
y_{1} y_{2} y_{7} y_{8})+ \\
&&\\
&&+ a(21) a(78) a(43) a(56) \delta_{18} \delta_{36} U_{4725}
(x_4 x_5 y_2 y_7 - x_2 x_7 y_4 y_5) \\
&&\\
B_{12;34}&=&U_{3214} (x_2 x_3 y_1 y_4 - x_1 x_4 y_2 y_3)
\end{array}
$$
where $a(ij)$ denotes the antisymmetrizer, i.e.
$a(ij)F_{ij} =F_{ij} -F_{ji}$.
The coefficients $B_{1234;56}$ and $B_{1234;5678}$ vanish.

In its full form the TSRPA is quite complicated for practical calculations.
It is possible to simplify the problem by using of the
boson operators $A^{+} , A $ ($b^{+}, b$) or the phonon operators $Q^{+} , Q$
constructed from $A^{+} , A $ ($b^{+}, b$)
as building blocks in the trial wave function (14) and the thermal Hamiltonian
(6), as was done in \cite{kosvdo} extending the quasiparticle-phonon
nuclear model~\cite{sol} to $T \neq 0$.

Partial support from RFBR under grant 96-15-96729 is acknowledged.

\end{document}